\documentclass[aps,prl,preprint,groupedaddress,showpacs]{revtex4}

\usepackage{graphicx}
\usepackage{amsmath}
\begin{document}



\title{Spin-flip processes and ultrafast magnetization dynamics in Co - unifying the microscopic and macroscopic view of femtosecond magnetism}


\author{M. Cinchetti}
\email[]{cinchett@rhrk.uni-kl.de}
\author{ M. S\'{a}nchez Albaneda}
\author{D.~Hoffmann}
\author{T. Roth}
\author{J.-P.~W\"{u}stenberg}
\author{M.~Krau{\ss}}
\author{ O. Andreyev}
\author{H. C. Schneider}
\author{M. Bauer}
\author{ M. Aeschlimann }
\affiliation{University of Kaiserslautern, Physics Department,
Erwin-Schr\"{o}dinger-Str. 46, 67663 Kaiserslautern, Germany}


\date{\today}

\begin{abstract}
The femtosecond magnetization dynamics of a thin cobalt film
excited with ultrashort laser pulses has been studied using two
complementary pump-probe techniques, namely  spin-, energy- and
time-resolved photoemission and time-resolved magneto-optical Kerr
effect. Combining the two methods it is possible to identify the
microscopic electron spin-flip mechanisms responsible for the
ultrafast macroscopic magnetization dynamics of the cobalt film.
In particular, we show that electron-magnon excitation does not
affect the overall magnetization even though it is an efficient
spin-flip channel on the sub-200\,fs timescale. Instead we find
experimental evidence for the relevance of Elliott-Yafet type
spin-flip processes for the ultrafast demagnetization taking place
on a time scale of 300\,fs.
\end{abstract}


\maketitle


Understanding the microscopic processes governing magnetization
dynamics on the femtosecond time scale is one of the most
challenging and interesting issues in today's condensed matter
physics.  There is nowadays much experimental evidence that the
excitation of a ferromagnetic film with an ultrashort laser pulse
gives rise to a demagnetization on a time scale of a few hundred
femtoseconds \cite{Beaurepaire96, Hohlfeld97,Scholl97, Güdde99,
Koopmans00, Rhie03}. However, the microscopic processes
responsible for these ultrafast changes in the macroscopic
magnetism of the ferromagnet have not been unambiguously
identified yet. A complete understanding of such processes would
be extremely helpful for technical applications like ultrafast
magnetization-switching schemes for read-write processes.  Most of
the recent experimental approaches applied in this field employed
time-resolved magneto-optical strategies
\cite{Beaurepaire96,Hohlfeld97, Güdde99, Koopmans00, Rhie03}. Such
experiments access the \textit{ensemble} spin dynamics taking
place on the femtosecond time scale.  On the other hand, spin- and
time-resolved photoemission addresses the dynamics of
\textit{single} electron spin-flip processes, which are
responsible - from a microscopic point of view - for the observed
changes in the sample magnetization. Bringing together these two
experimental methods has, therefore, the potential for an
efficient access to the microscopic view of the ultrafast dynamics
of macroscopic magnetism.

In this letter such a complementary approach is presented for the
first time and is used to obtain a more comprehensive
understanding of the ultrafast magnetization dynamics of a thin
cobalt film. Time-resolved magneto-optical Kerr effect (TR-MOKE),
as well as spin-, energy- and time-resolved two-photon
photoemission (SETR-2PPE) are used to address respectively the
macroscopic and microscopic dynamical response of the ferromagnet
induced by ultrashort optical excitation. SETR-2PPE probes the
time evolution of the spin polarization of single optically
excited electrons at a time-resolution down to a few femtoseconds.
In comparison with TR-MOKE data we find that characteristic
features in the observed complex time dependence of the SETR-2PPE
signal can be directly correlated to changes in the magnetization
of the sample. In this way, we are able to draw conclusions about
the microscopic spin-flip mechanisms leading to the observed
macroscopic ultrafast demagnetization of the thin Co film.

There are three main mechanisms through which an excited electron
can flip its spin in a ferromagnetic metal: (i) a Stoner
excitation (ii) an inelastic electron-spinwave scattering event
(magnon excitation), or (iii) an Elliott-Yafet type of spin-flip
scattering with impurities or phonons \cite{Elliot54,Yafet}.
Whereas for excitation energies above the vacuum energy
($E-E_F>\Phi$, $\Phi$=work function), channels (i) and (ii) have
been identified within static SPEELS measurements
\cite{Vollmer04,Etzkorn04}, there exists no direct experimental
access to optically excited electrons which are most relevant for
laser induced ultrafast demagnetization. Thus, the excitation
regime essential for ultrafast laser demagnetization
($E-E_F<\Phi$) remains to be explored experimentally.
Theoretically, short wavelength spinwave excitations have been
studied in \cite{Mills98,Mills04}. According to these studies,
spinwave scattering dominates Stoner excitation processes in the
very low energy regime (roughly $(E-E_{F})<3$\,eV). Moreover,
first-principles GW and T-matrix calculations including spin-flip
scattering \cite{Echenique04} have shown that the coupling to
spinwaves is an important inelastic decay channel of excited
electrons in itinerant ferromagnets, especially for energies
$(E-E_{F})<1.5$\,eV. On the other hand, a detailed analysis of
TR-MOKE experiments has provided some evidence that Elliott-Yafet
processes have to be considered for a correct interpretation of
femtosecond magnetization data \cite{KoopmansPRL05}.
 In the following we will show that
electron-magnon interaction indeed contributes significantly to
the spin dynamics of single electrons excited close to the Fermi
level ($E_{F}$) in a Co thin film. However we also find that this
process cannot account for the observed ultrafast demagnetization.
Instead we identify clear support for the Elliott-Yafet scenario.

The detailed experimental setup for SETR-2PPE has been described
elsewhere \cite{Knorren00}. Briefly, we use the second harmonic of
a femtosecond Ti:sapphire laser  with photon energy
$h\nu=3.1$\,eV. The beam is divided into two equally intense,
collinear and orthogonally polarized beams with an adjustable time
delay.  The pump photons (here, s-polarized) excite electrons from
their ground state into an unoccupied intermediate state. The
delayed probe photons (p-polarized) excite the electrons above the
vacuum level. The angle of incidence of the laser on the sample is
$45^{\circ}$; the fluence of the pump and probe pulses is
3$\mu$J/cm$^2$. The photoemitted electrons are analyzed in normal
emission geometry in an ultrahigh vacuum system equipped with a
commercial cylindrical sector energy analyzer (Focus CSA 300) and
a spin detector (Focus SPLEED).

Co films with a thickness of 10\,nm are grown  quasi-epitaxially
 on a Cu(001) substrate, and  possess  in-plane
magnetization easy axis along the (110) direction of the Cu
crystal. For the SETR-2PPE experiments the Co films are remanently
magnetized by an external field applied along the  easy axis
direction. The spin detector measures the projection of the spin
polarization along the  easy axis:
$P=(N^{+}-N^{-})/(N^{+}+N^{-})$, where $N^{+}(N^{-})$ is the
number of detected majority (minority) electrons.

After the SETR-2PPE experiments, the same Co film has been
investigated ex-situ by TR-MOKE. A two-color pump-probe experiment
has been used, where the pump pulses are generated by a multipass
amplifier system operating at a central wavelength of 800\,nm and
delivering 50\,fs pulses with $1.6$\,mJ/pulse at a repetition rate
of 1\,kHz. The probe pulses are obtained by frequency doubling a
small fraction ($\approx 1/10$) of the pump beam in a
$\beta$-barium-borate crystal. The bichromatic approach enables an
efficient suppression of bleaching effects which may otherwise
lead to a misinterpretation of the MOKE data \cite{Koopmans00}.
The experiments were performed by recording, for every pump-probe
delay, full hysteresis loops in longitudinal configuration. In
this way the shape of the hysteresis  can be used as a monitor of
the sample quality during measurements. The external field vector
is applied parallel to the in-plane easy axis direction. The inset
of Figure \ref{Fig3} shows two exemplary hysteresis loops recorded
respectively without the pump pulse (black curve) and 675\,fs
after it (grey curve).

\begin{figure}[h!]
\begin{center}
\includegraphics[width=8cm,keepaspectratio=true]{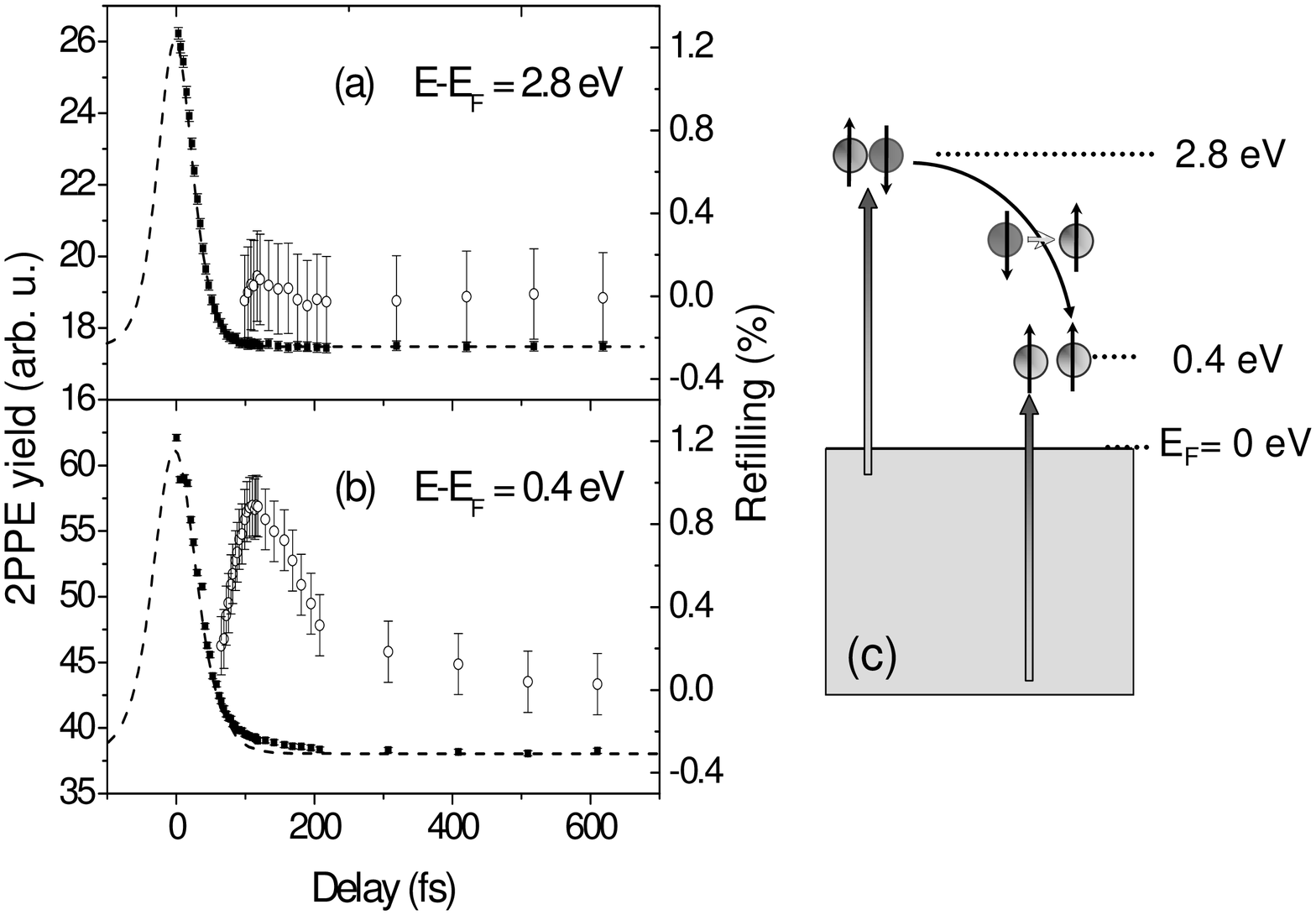}
\caption{M.\ Cinchetti, Physical Review Letters} \label{Fig1}
\end{center}
\end{figure}

In a time- and energy-resolved 2PPE measurement the recorded
quantity (cross-correlation curve) is the number of excited
electrons at a certain energy as a function of the time delay
between the pump and the probe pulse (population decay). Figure
\ref{Fig1} (a) and (b) show the cross-correlation curve of an
intermediate state with an energy of  2.8\,eV and 0.4\,eV above
$E_{F}$, respectively. In case the 2PPE signal is governed
exclusively by an exponential population decay of the intermediate
state, the experimental data can be fitted by a rate-equation
model, where the lifetime $\tau$ of the selected intermediate
level is the only fit parameter. Corresponding calculations to
simulate our experimental data were performed using the  formalism
described in \cite{Bauer99} in the limit of rapid dephasing
between the electronic excitation and the driving laser field. The
dashed lines in Figure \ref{Fig1} correspond to such a fit.
Whereas the data recorded at 2.8\,eV intermediate energy (a) are
almost perfectly described by this model, we find for the lower
energy (b) rather significant deviations  at temporal delays
between 50 and 200\,fs. In order to quantify these deviations, we
evaluate the refilling $R =(Y_{m}-Y_{f})/(Y_{m}+Y_{f})$, where
$Y_{m}$ is the measured 2PPE yield, and $Y_{f}$ the result of the
rate-equation model. The obtained curves are shown on the right
scale (open circles). Clearly, at 2.8\,eV intermediate energy $R$
is almost zero, while at 0.4\,eV it is significantly different
from zero up to 200\,fs delay, and reaches its maximum value of
$\sim 1$\,$\%$ at 120\,fs. The explanation for this additional
contribution to the 2PPE signal is that secondary electrons
scatter into the probed intermediate level and lead to a
repopulation (refilling) of the state itself (see Figure
\ref{Fig1} (c)). From the values of $R$ we can conclude that in
our case refilling electrons constitute at maximum $1$\,$\%$ of
the electron population at the considered intermediate level.
There are different relaxation mechanisms resulting in refilling
electrons; the most relevant is the scattering of an optically
excited (hot) electron in a higher excited state with a cold
electron from below $E_{F}$. Further details on refilling can be
found in \cite{Knorren00}.

\begin{figure}[h!]
\begin{center}
\includegraphics[width=8cm,keepaspectratio=true]{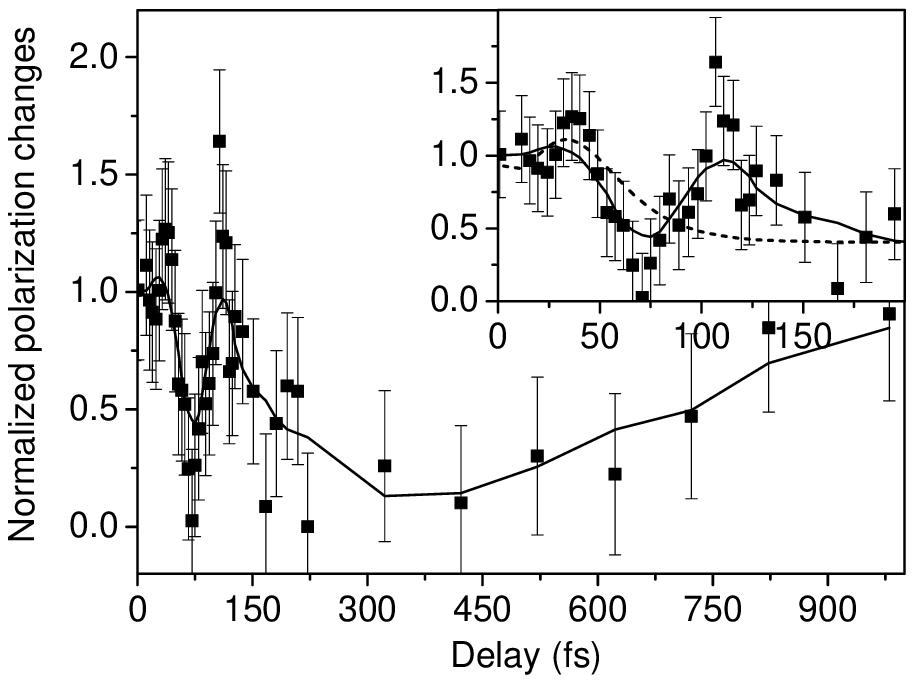}
\caption{M.\ Cinchetti, Physical Review Letters} \label{Fig2}
\end{center}
\end{figure}

\begin{figure}[h!]
\begin{center}
\includegraphics[width=8cm,keepaspectratio=true]{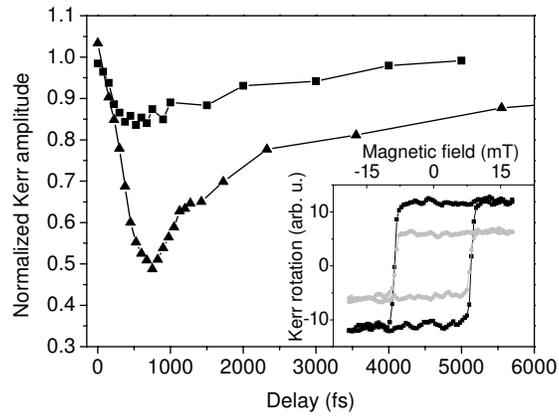}
\caption{M.\ Cinchetti, Physical Review Letters}
 \label{Fig3}
\end{center}
\end{figure}

Let us now concentrate on the time evolution of the spin
polarization of the refilling electrons in the intermediate state
located $0.4$\,eV above $E_{F}$ as detected by SETR-2PPE. Our aim
is to identify which kind of spin-flip scattering events take
place during the decay process as schematically shown in Figure
\ref{Fig1} (c). Figure \ref{Fig2} shows the dependence of the
in-plane polarization of electrons in the intermediate state at
this energy. In particular, we plot the normalized polarization
changes, defined as $\widetilde{\Delta
P}=(P-P_{\textmd{min}})/(P_0-P_{\textmd{min}})$, where $P_0$ is
the polarization at zero delay and $P_{\textmd{min}}$ the minimum
of the polarization over the studied delay range.
$\widetilde{\Delta P}$ exhibits a complex time-dependence with two
maxima at 30\,fs and 120\,fs, and two minima at 75\,fs and
350\,fs.

To separate different contributions to the spin- and
energy-resolved electron dynamics, we use a numerical model which
phenomenologically includes carrier Boltzmann scattering together
with carrier transitions due to pump and probe pulses. This model
 shows which features of the experimental results can
be explained  without the existence of magnetic correlations (such
as magnon excitation). More precisely, we solve dynamical,
energy-resolved equations for the electronic occupation $n_s(E)$
of single-particle states with spin $s$ and energy $E$ of the form

\begin{align}
\frac{\partial }{\partial t}n_{s}(E)= &-\Gamma
_{s}^{\mathrm{out}}(E)n_{s}(E)
+\Gamma _{s}^{\mathrm{in}}(E)\left[ 1-n_{s}(E)\right] \nonumber \\
&+  \left.\frac{\partial }{\partial
t}n_{s}(E)\right|_{\mathrm{pp}} \label{rate-general}
\end{align}
where the last term on the right-hand side describes electronic
transitions due to the  pump and probe laser fields and includes
the density of states which is taken from experiment. The in and
out-scattering rates in Eq.~(\ref{rate-general}) for weak
nonequilibrium situations are determined from the measured
spin-dependent electron lifetimes using $ \Gamma
_{s}^{\mathrm{out}}(E) = -b(-E)/\tau_s(E)$, where $b(E) =
[\exp(\beta E)-1]^{-1}$ is the Bose function, together with the
detailed-balance relation $ \Gamma _{s}^{\mathrm{out}}(E) =
\exp(\beta E) \Gamma _{s}^{\mathrm{in}}(E)$. Using these relations
implies that the dynamics obtained from Eq.~(\ref{rate-general})
describes transitions involving scattering events with both
minority and majority electrons, but does not account for the full
correlated electron dynamics. In particular, our approach can
describe the influence of the dynamical single-particle occupation
on the experiment, which includes effects such as cascade
electronic scattering events.

The result of the simulation for delays up to 200\,fs is shown in
the inset of Figure \ref{Fig2} (dashed line), together with the
measured data. Clearly, a good agreement between experiment and
calculations is achieved only for the first 100\,fs. Here, the
spin polarization first increases from its value at zero delay to
a maximum close to 30\,fs. This is due to the longer lifetime of
majority electrons compared to the lifetime of minority electrons
\cite{Aeschlimann97}. This difference implies that minority
electrons depopulate the intermediate state at a faster rate than
majority electrons, leading to a transient increase in the spin
polarization (spin-filter effect).  Between 30 and 75\,fs the
spin-filter effect starts to become irrelevant and the
polarization decreases to the background value given by the
spin-polarization excited by the pump and probe pulse separately.
The successive increase of the spin polarization in the
experiment, with a maximum around 120\,fs, is not reproduced by
the simulation. At this time delay we have a maximum of the
refilling curve $R$ in Figure \ref{Fig1} (b). In order to
interpret this behavior correctly, let us first observe that the
calculations in \cite{Echenique04} predict that minority electrons
exhibit a higher cross-section for magnon excitation compared to
majority electrons. Combining this results with the fact that our
numerical simulation does not include magnon excitation, we can
conclude that this process causes the measured increase of the
spin polarization around 120\,fs. In other words, the intermediate
state at $0.4$\,eV above $E_F$ is refilled by secondary electrons
originating from energetically higher-lying minority electrons
(with energy up to $3.1$\,eV above $E_F$): those electrons have a
higher probability than majority electrons to scatter
inelastically with a magnon and flip thereby their spin, turning
into majority electrons and thus causing an increase in the
measured spin polarization (see Figure \ref{Fig1} (c)).

We now discuss the behavior of the spin polarization for time
delays between approximately 150 and 1000\,fs. It shows a drop
with a minimum around 350\,fs and a successive recovery. For such
time delays, the number of pump laser excited electrons is so
small that the observed behavior must be due to a change in the
magnetic state of the electrons within the Fermi see. Indeed,
after optical excitation the excited electrons and the produced
secondary electrons thermalize to a hot Fermi-Dirac distribution
on such a time scale. This implies that the feature observed in
Figure \ref{Fig2} is related to a process in which the total
magnetization is changing, and should be observed also with
TR-MOKE. Figure \ref{Fig3} shows the TR-MOKE signal recorded for
10\,mJ/cm$^{2}$ (triangles) and 2\,mJ/cm$^{2}$ (squares) pump
fluence. The curves are obtained by plotting, for a given
pump-probe delay, the Kerr amplitude of the corresponding
hysteresis loop, normalized by the value at zero delay. They show
an ultrafast drop followed by a recovery. By reducing the pump
fluence the minimum of the curve shifts to lower time delays. For
the lower fluence the minimum is reached at approximately 400\,fs,
close to the minimum observed in the SETR-2PPE experiment. This
fact gives a clear indication that the mechanism responsible for
the reduction of the polarization in the SETR-2PPE experiment is
the same one causing a reduction of the TR-MOKE signal, i.e.\ it
is a process in which the total magnetization is changing.
Koopmans \textit{et al.} \cite{KoopmannsJMMM05,KoopmansPRL05} have
argued
 that the ultrafast
demagnetization observed in TR-MOKE experiments can be
microscopically explained by assuming Elliott-Yafet type
 of spin-flip scattering of electrons with impurities or phonons.
According to the same authors, such process leads to a decrease of
the minimum of the TR-MOKE curves with the pump fluence. Such a
behavior was indeed observed in \cite{vanKampen03}, though, at
lower fluencies and thereby with more subtle variations. Our
TR-MOKE results give a valuable support to such interpretation and
moreover, through the comparison with SETR-2PPE  provide the
experimental proof of the dominating role of Elliott-Yafet
scattering in a microscopic description of ultrafast
demagnetization.

In conclusion we have successfully combined two experimental
techniques, one (TR-MOKE) giving an insight into the ensemble
femtosecond spin dynamics after laser excitation, the other
(SETR-2PPE) providing additional information on the microscopic
spin-flip processes responsible for the observed magnetization
dynamics. We were able to conclude that even if microscopically
spinwave excitation plays an important role in the ultrafast spin
dependent electron dynamics, the ultrafast demagnetization of the
studied thin Co film is mainly due to Elliott-Yafet type spin-flip
processes.

\begin{acknowledgments}
Financial support from DFG priority program 1133 and the European
research training network HPRN-CT-2002-00318 ULTRASWITCH is
gratefully acknowledged. We thank F.\ Steeb for his support with
the femtosecond laser system. One of us (M.C.) would like to thank
R.\ Santachiara for stimulating discussions.
\end{acknowledgments}


\newpage

\section{figures}

\newpage

\newpage

\newpage

\section{Captions}

\begin{description}
\item{FIG.\ 1:} (a) and (b). Left scale: 2PPE yield from an
intermediate state with energy of $2.8$\,eV (a) and $0.4$\,eV (b)
as a function of the delay time between the pump and the probe
pulse. The fit of the experimental curves with a rate-equation
model is plotted with  dashed lines.  Right scale: Refilling
parameter $R$ as derived from the measured
curves and the fitting curves.\\
(c) Schematic picture of a possible spin-dependent refilling
process of the intermediate state at $0.4$\,eV, as discussed in
the text.

\item{FIG.\ 2:} Time dependence of the spin polarization of
electrons in the intermediate state at $0.4$\,eV above $E_F$ as
detected by SETR-2PPE. Inset: a zoomed picture of the same curve
for time delays between $0$\,fs and $200$\,fs and the result of
the numerical model discussed in the text (dashed line). The
continuous line (obtained by smoothing the original data) is
depicted as a guide to the eye for the reader.

\item{FIG.\ 3:} Results of TR-MOKE  measurements on the same Co
sample of Figure \ref{Fig2}, taken with a pump fluence of
10\,mJ/cm$^{2}$ (triangles) and 2\,mJ/cm$^{2}$ (squares). Inset:
two exemplary hysteresis loops recorded respectively without the
pump pulse (black curve) and 675\,fs after it (grey curve). The
pump power was 10\,mJ/cm$^{2}$.
\end{description}
\end{document}